\documentclass{aa}

\usepackage{graphicx}
\usepackage{multirow}
\usepackage{natbib}
\usepackage{booktabs}

\usepackage{threeparttable}
\usepackage{hyperref}
\usepackage{graphicx,float,amsmath,bm,epsfig,epstopdf}
\usepackage{txfonts}
\usepackage{threeparttable}

\begin{document}

   \title{Magnetic interaction analysis of multiple interplanetary coronal mass ejections leading to a historic geomagnetic storm in May 2024}

   \author{Sanchita Pal\inst{\ref{inst1}}\and Cecilia Mac Cormack \inst{\ref{inst2},\ref{inst3}}\and Emilia~K.~J.~Kilpua \inst{\ref{inst4}} \and Yogesh \inst{\ref{inst2},\ref{inst3}} \and Lan K. Jian \inst{\ref{inst2}} \and Teresa Nieves-Chinchilla \inst{\ref{inst2}} }


\institute{NASA Postdoctoral Program Fellow, NASA Goddard Space Flight Center, Greenbelt, MD 20771, USA \label{inst1} \and Heliophysics Science Division, NASA Goddard Space Flight Center, Greenbelt, MD 20771, USA \label{inst2} \and The Catholic University of America, Washington, DC 20064, USA \label{inst3}
\and Department of Physics, University of Helsinki, P.O. Box 64, FI-00014 Helsinki, Finland \label{inst4}}

 \abstract
 {Interplanetary coronal mass ejections (ICMEs), the large-scale eruptive phenomena capable of shedding a huge amount of solar magnetic helicity and energy are potential in driving strong geomagnetic storms. They complexly evolve while preceded and followed by other large-scale structures e.g. ICMEs. Magnetic interaction among multiple ICMEs may result intense and long-lived geomagnetic storms.  }
 {Our aim is to understand the reason of substantial changes in the geoeffectivity of two meso-scale separated counterparts of a complex solar wind structure through investigating their magnetic content e.g. helicity, energy and magnetic interaction among multiple ICMEs.}
 {We utilized the in situ observations of solar wind from Wind and Solar Terrestrial Relations Observatory-A (STA) spacecraft during the strongest geomagnetic storm period in past two decades on May 10-11, 2024. We performed heliospheric imaging analysis to locate the solar sources, investigate the interplanetary propagation and Earth-arrival of the driver, and performed a time-frequency domain analysis of the in situ magnetic field vectors in injection and inertial ranges of magneto-hydrodynamic (MHD) turbulence to quantify the driver's magnetic content at two counterparts. }
 {Our investigation confirms complex interactions among five ICMEs resulting in distinct counterparts within a coalescing large-scale structure. These counterparts possess substantially different magnetic contents. }
 {We conclude that the complex counterpart resulted from the interaction among common-origin ICMEs observed by STA, favorably orientated for magnetic reconnection, had 1.6 and 2.8 times higher total magnetic energy and helicity, respectively, than the counterpart containing a left-handed filament-origin ICME observed by Wind. The left-handed ICME non-favorably oriented for magnetic reconnection with the surrounding right-handed, common-origin ICMEs. Therefore, two medium-separated counterparts despite belonging to a common solar wind structure, were potential to lead different geoeffectivity. This ultimately challenges space weather predictions based on early observations.}

\maketitle

\section{Introduction} \label{sec:1}
Interplanetary coronal mass ejections \citep[ICMEs, e.g.][]{ 2017LRSP...14....5K} are the heliospheric manifestation of large-scale eruptive solar phenomena that impulsively release excess magnetic energy stored in twisted solar coronal field lines \citep[e.g.][]{2017ApJ...851..123P}. As they depart the corona and propagate through the heliosphere, they experience complex interactions with the surrounding plasma and magnetic fields that may lead to their significant evolution and distortion of their structure \citep[e.g.][]{2024ApJ...975..169W, 2023FrASS..1095805P, 2022ApJ...938...13B}, deceleration and/or acceleration \citep[e.g.][]{2012JGRA..11711101S}, reconfiguration,  magnetic flux erosion \citep[e.g.][]{2022FrASS...9.3676P, 2021A&A...650A.176P, 2020GeoRL..4786372P}, and accumulation of magnetic field lines \citep[e.g.][]{2017ApJ...837L..17T, 2017LRSP...14....5K}. Understanding physical processes driving solar magnetic field, their emergence and variability, their manifestation as eruptive events that causally connect the Sun–geospace environment and processes leading complex evolution of large-scale eruptions in the heliosphere have been reviewed by \cite{2023JASTP.24806081N}, \cite{2017SSRv..212.1159M} and \cite{2020SoPh..295...61L}.

\par

Studying interacting ICMEs attracts interests as they provide an ideal opportunity to study the complexity of space plasmas, such as propagation of shock wave within a magnetically dominated structure \citep{2005ApJ...634..651L, 2009AnGeo..27.3479L, 2023FrASS..1095805P}, and magnetic and non-magnetic interactions among large-scale structures \citep{2004AnGeo..22.2245S, 2021A&A...656A...5T, 2020ApJ...905L..12T}. Furthermore, interacting ICMEs may result in intense and long-lived geomagnetic storms \citep{2006JGRA..11111104F, 2006AdSpR..38..498F, 2006JGRA..111.1103X,2014NatCo...5.3481L} by embedding extended periods of strong southward-directed magnetic field vector $B_z$ \citep{ 2003GeoRL..30.1700W}. By numerical simulations, \cite{2004AnGeo..22.2245S} explained the different nature of interactions among ICMEs e.g. magnetic interaction and non-elastic collision. An equalization of speeds is achieved in all interacting ICMEs, regardless of the type of interactions. A magnetic interaction between ICMEs depends on their relative trajectories and sense of magnetic field topology. When the two ICMEs have parallel axes with the same sense of magnetic field rotation, magnetic reconnection occurs at their interface, resulting into a single complex magnetic structure  \citep{2013ApJ...778...20L}. However, if field line characteristics do not allow reconnection, ICMEs collide with highly non-elastic manner. They distort around each other due to compressibility nature of their magnetized plasma.  
\par
In general, ICMEs embed highly twisted magnetic field lines \citep{2023FrASS..1014838N} and can be unambiguously identified in the in situ data as regions of high magnetic helicity ($H_m$) and energy ($E_m$). The $H_m$ is a measure of the field line twist or writhe \citep{1978mfge.book.....M} and can not be achieved without the information of topological (spatial) properties of field lines. \cite{2012ApJ...751...19T} advanced a novel technique provided by \cite{1982JGR....87.6011M} to compute the reduced $H_m$ from the Fourier transform of the magnetic field auto-correlation matrix. While the trace of the symmetric part of the spectral matrix measures the $E_m$, the imaginary part of the matrix measures the $H_m$ \citep{1953tht..book.....B, 1981PhFl...24..825M, 1982JGR....87.6011M}.
This advanced technique has been widely used in recent studies \citep[e.g.][]{2013ApJ...776....3T, 2016ApJ...826..205T, 2019ApJ...885..120T, 2020ApJS..246...26Z, 2021A&A...650A..12Z, 2020ApJ...905L..12T, 2023ApJ...952..111T, 2022A&A...664L...8A} to identify and investigate the MHD properties of helical field lines in the solar wind. 
\par
The complexity of the magnetic field structure within ICME ejecta varies significantly from case to case and therefore, their identification  may be challenging \citep{2025SSRv..221...12A}. At Earth, they reach a radial extension of about 0.25 au \citep{1982JGR....87..613K} with a timescale ranging up to 64 hours \citep{2019ApJ...885..120T}. Although a southward magnetic vector within ICMEs is essential to drive a geomagnetic storm, this is not a sufficient condition as it is also necessary that the ICME carries a considerable amount of energy (kinetic or magnetic) to ensure an efficient energy release during the storm. An automated scheme to detect ICMEs in solar wind and assess their geoeffectiveness using helicity and energy spectrum analysis has been implemented successfully \citep{2019ApJ...885..120T}. A greater amount of available ICME magnetic energy during its reconnection with the geomagnetic field leads more efficient energy release into the magnetosphere and transform to kinetic and thermal energy that further results in acceleration of particle and increasing density in the Earth's upper atmosphere. This may result in serious disruptions in satellite signal propagation, radio communication, positioning systems and additional drag to satellites changing their orbits \citep{2024SpWea..2203716B,10.3389/fspas.2025.1572313}. Therefore, immense efforts have been employed to reliably detect and predict ICME-driven space weather using data-driven magnetohydrodynamics (MHD) simulations \citep[e.g.][]{2023FrASS..1005797Z}, hybrid empirical/physics-based MHD model \citep[e.g.][]{2008SpWea...6.8001O, 2024ApJ...976..126M}, data-constrained analytical methods \citep[e.g.][]{2022SpWea..2002914K, 2020ApJ...888..121S, 2022A&A...665A.110P, 2021ApJ...920...65P}, and physics-based artificial intelligence techniques \citep[e.g.][]{2024ApJ...972...94P, 2025arXiv250509365R, 2024SoPh..299..131N}. 
\par
The recent May 2024 storm was one of the largest geomagnetic storms that have happened since the beginning of the space age in 1957  with the disturbance storm time (Dst) index reaching at a minimum of -412 nT at 02:00 UTC on May 11 \citep{2025NatSR..15.5922D}.  In total, only 11 great geomagnetic storms with Dst $\le$ 350 nT \citep{2011JASTP..73.1447G} has occurred since start of the space era. The associated structured solar wind was probed simultaneously by spacecraft at Lagrangian(L)-1 and the Solar Terrestrial Relations Observatory-A (STA), which was $ 12.6^\circ$ westward and $ 1.5^\circ$ northward from the Sun-Earth line. The impacts of this storm on Earth's upper atmosphere and geomagnetically induced current (GIC) have been extensively studied \citep[e.g.][]{2025SpWea..2304245Z, 2025SpWea..2304235C, 2025SpWea..2304197Z, 2025JGRA..13033601A, 2025JGRA..13033627W, 2025GeoRL..5215104Z, 2024SpWea..2204126T}. Recent studies \citep[e.g.][]{2024ApJ...974L...8L, 2024A&A...692A.112W, 2025ApJ...979...49H, 2025ApJ...981...76T, 2025SpWea..2304260W, 2024JASS...41..171K, khuntia2025evolution} provided detailed analysis of event-related solar origins, connections between solar and interplanetary counterparts, thermodynamics and evolution of CME-CME interactions, and simulated ICME propagation and their arrival. Using two empirical formulas provided by \cite{2000JGR...105.7707O} and \cite{1975JGR....80.4204B}, \cite{2024ApJ...974L...8L} estimated that the ICME structure observed at STA would have caused the  minimum $Dst$ ($Dst_{min}$) that was almost $\sim 116$ nT lower than observed at Earth. Furthermore, their estimated $Dst_{min}$ was 8\% smaller than the observed $Dst_{min}$ at Earth.
\par

A complex structured solar wind may appear with substantial differences in multipoint observations with meso-scale separations \citep{2023FrASS..1095805P, 2025SpWea..2304189L, 2024ApJ...963..108P}, which lead to difficulties in their reliable predictions. In this work, we utilise the occasion of multipoint meso-scale separated probing of a very complex solar wind structure and derive differences in their magnetic content potential to drive geoeffectivity at two probing locations. We address the reason of such differences using multiple ICME-ICME interactions phenomena. Furthermore, we attempt to explain why two moderately separated counterparts belonging to a highly structured solar wind exhibit significantly different geoeffectiveness. In Section \ref{sec2}, we provide details of data and method of our analysis. Section \ref{sec3} presents results and analyzes them. Finally, in Section \ref{sec4}, we discuss and conclude our study. \par

 \par
\section{Data and Methodology} \label{sec2}
In this study, we utilise multipoint in situ observations from Magnetic Field Investigation \citep[MFI;][]{1995SSRv...71..207L}, Solar Wind Experiment \citep[SWE;][] {1995SSRv...71...55O} and 3D Plasma Analyzer \citep[3DP;][]{1995SSRv...71..125L} instruments onboard Wind and Magnetometer (MAG), Solar Wind Electron Analyser (SWEA) of In situ Measurements of Particles and CME Transients (IMPACT) suite \citep{2008SSRv..136..203A}, and Plasma and Suprathermal Ion Composition (PLASTIC) instrument \citep{2008SSRv..136..437G} onboard STA. We use remote sensing observations from heliospheric imagers (HI1 and HI2) and COR2 of the Sun-Earth Connection Coronal and Heliospheric Investigation \citep[SECCHI:][]{howard_2008} suite onboard STA. We obtain science-level in situ data from NASA's Coordinated Data Analysis Web (CDAWeb) with the highest available resolution. As Level 2 IMPACT/MAG data had significant data gap, we decided to use Level 1 MAG data with the highest resolution (8 samples/sec) in this work.\par
We follow the method of obtaining solar wind magnetic helicity \citep{2012ApJ...751...19T},
\begin{equation}
    H_m(k,t)=2\Im[W_y^*(k,t). W_z(k,t)]/k,
\end{equation}
and normalized helicity \citep{2013ApJ...776....3T}
\begin{equation}
    \sigma_m(k,t)=\frac{2\Im[W_y^*(k,t). W_z(k,t)]}{|W_x{k,t}|^2+|W_y{k,t}|^2+|W_z{k,t}|^2},
\end{equation}
where $W_x$,  $W_y$ and $W_z$ represent the Morlet wavelet transform of $x_{gse}$, $y_{gse}$, and $z_{gse}$ magnetic field components in geocentric solar ecliptic (GSE) coordinate system, and $k$ is the wave number. This tool extends a spectrum-based formula of helicity derivation by \cite{1982JGR....87.6011M} to the time domain, allow proper localization of magnetic helicity structures in solar wind and has been extensively used to study MHD properties of ICMEs \citep{2013ApJ...776....3T, 2016ApJ...826..205T, 2019ApJ...885..120T, 2020ApJ...905L..12T, 2021A&A...656A...5T}. While $H_m(k,t)\sim 0$ indicates untwisted field lines, $H_m(k,t)>0$ $(<0)$ corresponds to counter-clockwise (clockwise) field line rotation in outward magnetic sector and $H_m(k,t)<0$ $(>0)$ corresponds to counter-clockwise (clockwise) field line rotation in inward magnetic sector. An ejecta of intrinsic right-handed (left-handed) helicity is observed to have a positive (negative) value of $H_m (k,t)$, which indicates a counter-clockwise (clockwise) rotation of field lines in an outward magnetic sector in the heliosphere.   \par

Utilizing Morlet wavelet transform ($W$) of $\mathbf{B}$, we derive magnetic energy spectrum $E_m(k,t)$ of solar wind following \cite{2019ApJ...885..120T},
\begin{equation}
    E_m(k,t)=0.5\sum_{i=x_{gse},y_{gse},z_{gse}} W_i(k,t).
\end{equation}
Thus, the total magnetic energy in solar wind can be calculated by $E_M (t)=\int{k^{5/3}E_m(k,t) dk}$. Here, by multiplying $k^{5/3}$, we compensate $E_m(k,t)$ to highlight small-scale high-energy structures in solar wind because magnetic field fluctuations exhibit a 5/3 Kolmogorov-like spectrum \citep{2019ApJ...885..120T}. Similarly while estimating the total helicity $H_m(t)=\int k^{8/3}|H_m(k,t)|$, we multiplied 8/3 to $H_m(k,t)$ to highlight the small-scale helicity structures in helicity spectrum as it drops off with $k^{-8/3}$ \citep{1982JGR....87.6011M, 1986AnGeo...4...17B}. Values of $|H_m|$ and $E_m$ greater than the two standard deviation of the mean unperturbed solar wind condition are considered as threshold to identify ICMEs in solar wind \citep{2019ApJ...885..120T}.\par

To correlate in situ detections with their associated CMEs, we predict the arrival time of these structures using STEREO/SECCHI data and the harmonic mean (HM) method. This method assumes a circular geometry for the CME front, anchored to the Sun, and can be used to compute velocity, propagation angle, and arrival time of an ICME, \citep[see e.g.,][]{lugaz_2010,mostl_2011}. We implement this fitting using the Solar Angle-Time plot \href{https://solar.jhuapl.edu/Data-Products/COR2-HI.php}{(SATPLOT)} software tool, which generates a plot of elongation angle versus time (J-maps) from SECCHI COR2 and HI1 and HI2 data to track solar wind heliospheric structures and subsequently apply a fitting method to them to predict their arrival and impact on spacecraft in the heliosphere. The SATPLOT is available in the SECCHI software tree of SolarSoft \citep{1998SoPh..182..497F}.

\section{Results and Analysis}\label{sec3}
\subsection{Event Overview}
Several studies have attempted to connect ICME ejecta observed in situ during May 10-11, 2024, to their solar origins. A total of six CMEs have been identified  with suitable speeds and source positions to potentially encounter the Earth and/or STEREO-A. The details of these CMEs,  taken from the studies by \cite{2024ApJ...974L...8L}(L) and \cite{khuntia2025evolution}(K) using the graduated cylindrical shell technique \citep[GCS,][]{thernisien_2006, thernisien_2011}, are provided in Columns 3-9 of Table \ref{T1}. An association has been found with two M and three X-class flares originating from the AR13664 that was a super active region exhibiting extreme activity by producing 23 X-class and multiple M-class flares \citep{2025ApJ...979...31J} and a quiet-Sun left-handed filament \citep{2025SpWea..2304260W, 1998SoPh..182..107M} located to the south of the AR13667. In the `Associated CMEs' column of Table \ref{T1}, except for the propagation direction, we notice a significant disagreement in CME physical parameters between different studies. This could be related to  the separation between the SOHO and STA spacecraft being  far from the ideal to perform reliably the multi-spacecraft GCS fitting \citep{2012SoPh..281..167B}.\par

\begin{table*}[h]
\small
    \centering
    \begin{tabular}{|c|c|c|c|c|c|c|c|c|c|c|}
    \hline
    \multicolumn{2}{|c|}{\textbf{Complex solar wind}} & \multicolumn{7}{c|}{\textbf{Associated CMEs}} \\ 
    \multicolumn{2}{|c|}{\textbf{intervals}} & \multicolumn{7}{c|}{} \\ \hline
    \textbf{STA} & \textbf{Wind} & \textbf{Event/Source} & \textbf{Time (UT)} & \textbf{Dir} & \textbf{Speed (km/s)} & \textbf{A } & \textbf{AW ($^\circ$)} & \textbf{Tilt ($^\circ$)} \\ \hline
    \multirow{6}{*}{\textbf{S1}} & \textbf{W1} & CME1 (X1.0) &05/08, 5:36 &S14W09 (L) &750 (L) &0.96(L) &38 (L)  &-63 (L) \\ 
    & Start: 05/10, 19:00& AR13664 & &S08W16 (K) &967 (K)&0.27 (K) &48 (K)  &84 (K) \\ \cline{3-9} 
     & End: 05/11, 01:30& CME2 (M8.7) &05/08, 12:24 &S10W05 (L) &850 (L) &0.95 (L) &30(L)&-55 (L) \\  
     &  & AR13664  & &S16W13 (K) &1142 (K) &0.34 (K) &46(K)&27 (K) \\ \cline{2-9}
     & \textbf{W2} & CME3 (Fil) & 05/08, 19:12&N04E27 (K) &991 (K) &0.24 (K) & 30 (K)& 79 (K) \\ 
     Start: 05/10, 15:40& Start: 05/11, 02:48& AR13667&&&&&& \\ \cline{3-9} 
     End: 05/11, 11:20& End: 05/11, 09:10  & CME4 (X1.0) &05/08, 22:36& S16W07 (L)& 1480 (L)&0.72(L) &18 (L) &-10 (L) \\ 
      &  &AR13664  &&  S18W06 (K)& 1406 (K)&0.26(K) &36 (K) &15 (K) \\ \cline{3-9}
     &   & CME5 (M9.8) &05/08, 22:36& S15W38 (K)&1103 (K) &0.15 (K) & 32 (K)&-83 (K) \\
     & &AR13664 &&&& & & \\ \cline{2-9} 
     &   Start: 05/11, 11:30 & CME6 (X2.2) &05/09, 9:24 &S12W23 (L) &1480 (L) &0.84 (L) &16 (L) &-25 (L) \\ 
     &   End: 05/11, 17:00 &AR13664  &&S14W27 (K) &1746 (K) &0.28 (K) &46 (K) &-77 (K) \\ \hline
    \end{tabular}
    \caption{Helicity localized regions $W1, W2$ and $S1$ within the complex solar wind intervals observed by Wind and STA and associated CMEs with their source active regions (\textbf{Event/Source}), first arrival at coronagraph (\textbf{Time}), propagation direction (\textbf{Dir}), \textbf{Speed}, aspect ratio (\textbf{A}), angular width (\textbf{AW}) and \textbf{Tilt} measured close to the origin ($<30 Rs$) reported by previous studies \cite{2024ApJ...974L...8L}(L) and \cite{khuntia2025evolution} (K).}
    \label{T1}
\end{table*}
\par
\begin{figure*}[t]
  \centering
    \includegraphics[width=1\textwidth]{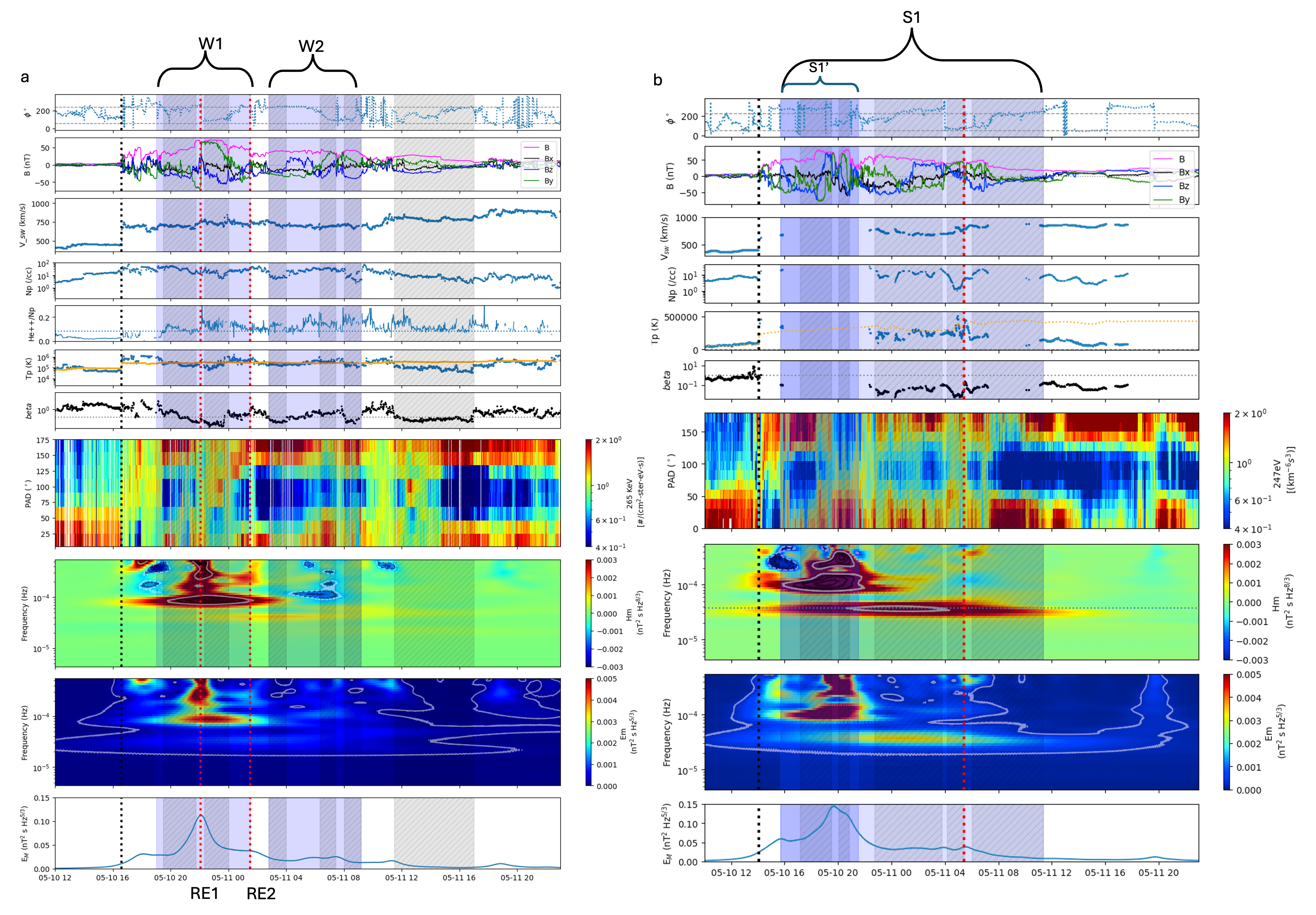}
  \caption{In situ observation during May 10-11, 2024 using (a) MAG, SWE instruments onboard Wind and (b) MAG, SWEA and PLASTIC instruments onboard STA spacecraft. The dotted lines in the $\phi$ panel indicate the nominal sector boundaries of the interplanetary magnetic field derived using a Parker spiral angle obtained from the local solar wind observations at Wind and STA. Here, $\phi$ values between (outside) the lines indicate the magnetic field in the away (toward) sector of the heliosphere. The blue shaded regions $W1$, $W2$, and $S1$ present the localized helicity regions in both time and frequency domains (see Section \ref{3.1} for details). The hatched intervals indicate an unperturbed core of ejecta, as discussed in Section \ref{3.2}. The 8th (7th) column in panel a (b) shows the energy flux (phase space density) of suprathermal electrons in 265 (247) eV energy bin.
  }
  \label{fig1}
\end{figure*}

In Figure \ref{fig1}, we provide in situ observations of solar wind magnetic and plasma properties, including magnetic field longitude angle $\phi$, magnetic field intensity $B$ and vectors $B_x, B_y, B_z$ in GSE coordinates, bulk speed $V_{sw}$, proton number density $N_p$, alpha to proton number density ratio $N_a/N_p$, proton temperature $T_p$ with expected ambient temperature $T_{ex}$ (shown using a yellow line over-plotted on the $T_p$ panel) derived from the correlation between $V_{sw}$ and $T_p$ \citep{1987JGR....9211189L}., $\beta$ (ratio of plasma thermal pressure to the magnetic pressure), suprathermal electron pitch angle distribution (PAD), $H_m$ and $E_m$ spectra and total magnetic energy derived from Wind and STA covering the peak activity solar wind period during May 10, 12:00 UT - May 11, 23:00 UT. While Wind had an uninterrupted plasma and magnetic data coverage during the investigated solar wind period, STA had many plasma data gaps. To perform a magnetic analysis of solar wind during the peak activity period, we analyze $H_m (k,t)$ and $E_m(k,t)$ in the injection frequency range that corresponds to the larger scale-size structures in solar wind. We investigate the localization of helicity in solar wind during May 10-11, 2024 in both Wind and STA observations in a timescale of 0.5-64 hours corresponding to frequency domain $\sim4.6e^{-6}-\sim5.5e^{-4}$ Hz. We provide the $H_m$ and $E_m$ spectra in Figure \ref{fig1}a and b obtained from Wind and STA, respectively, where the white contours enclose the highly helical regions having $|H_m|\ge 1/e |H_{m}|_{max}$. In the $E_m$ spectrum, the white contour encloses regions corresponding to $Em \ge2 \sigma \bar{E_{m,u}}$, where $\bar{E_{m,u}}$ is the mean of $E_m$ of unperturbed solar wind. The contours in $H_m$ and scale size corresponding to the highest $H_m$ value help us to assess the time extent and scale sizes $\Delta S$ of ejecta cores with maximum winding of field lines,  respectively. \par
\subsection{Complex solar wind intervals detected by Wind and STA}\label{3.1}
During the majority of the shown intervals the magnetic helicity is close to zero. The Wind spacecraft however observed localized patches of positive and negative helicity with $|H_m|$ being two standard deviation higher than the average $|H_m|$ of the unperturbed solar wind. The helicity patches were in a mostly outward magnetic field sector, revealing counter-clockwise and clockwise rotations with the size $\Delta S$ of the characteristic scale corresponding to the strong winding of magnetic field lines around its axis equal 3.2 hrs and 2.4 hrs and average $\sigma_m=0.8\pm 0.07$ and $-0.71\pm 0.11$, respectively. The blue shaded intervals in Figure \ref{fig1} indicated by $W1$ and $W2$ represent their temporal extensions. The $W1$ contains comparatively strongly wrapping $\mathbf{B}$ with maximum mean helicity, multiple intervals with lower $N_p$, $T_p<T_{ex}$, $\beta<0.1$, and $N_a/N_p$ peaks representing presence of multiple ejecta, corresponding to the highest $E_M(t)$ value during May 10-15, 2024. Region $W2$ features comparatively loosely wrapping left-handedly rotating field lines with unambiguously identified three intervals of $T_p<T_{ex}$ and $\beta<0.1$ and a couple of intervals with lower $N_p$. The region corresponding helicity and energy are less than that of $W1$. 
\par

At 12.6$^\circ$ West and 1.5$^\circ$ North from the Wind location (L1), STA observed different features in $H_m$ spectrum than those observed by Wind. Here, a tightly wound magnetic field forms a complex ejecta core with an average $\sigma_m=0.9\pm 0.03$ and a scale size of $\Delta S=7.4$ hrs spanned over a time range indicated by the shaded region $S1$ in Figure \ref{fig1}b. The $S1$ contains a complex ejecta with intervals of bi-directional PAD, lower $N_p$, $\beta$<0.1, and $T_p<T_{ex}$. Plasma data observed by STA had a significant data gap. Therefore, $T_{ex}$ for STA has been computed using shifted plasma data obtained from Wind \citep{2024ApJ...974L...8L}. 
\par
In a comparatively smaller frequency range ($\sim7e^{-5}-\sim2e^{-4}$ Hz), within $S1$, we observe another highly helical patch with counter-clockwise field line rotations extending in the region $S1'\subset S1$ (shown using darker blue interval in Figure \ref{fig1}b) having $\Delta S=3$ hrs equivalent to the $\Delta S$ of the localized positive helicity region $W1$ in the Wind observation. We noted that, the $S1'$ is preceded by another helical structure located in comparatively smaller scale, observed having a negative helicity sign in toward magnetic sector. This structure was also observed at Wind with a much lower helicity. The positive helical structure protruding to the smaller-scale size in front of the negative helicity patch in STA, likely resulted from the magnetic erosion of the outer layer of the preceding structure by the following one. This further implies that the two structures are engaged in magnetic reconnection. The interval $S1$ contains highly helical field lines and a significant amount of magnetic energy that is evident from the high $E_m$ values localized at a scale size equivalent to $S1$. Magnetic energy within it reaches its peak during $S1'$. We found that during the interval of May 10-11, the STA counterpart had a peak magnetic energy almost twice that of the Wind counterpart, although at both observations the energy curve had a quite similar shape.

\subsection{Ejecta cores within complex intervals} \label{3.2}
\begin{figure*}[t]
  \centering
    \includegraphics[width=1\textwidth]{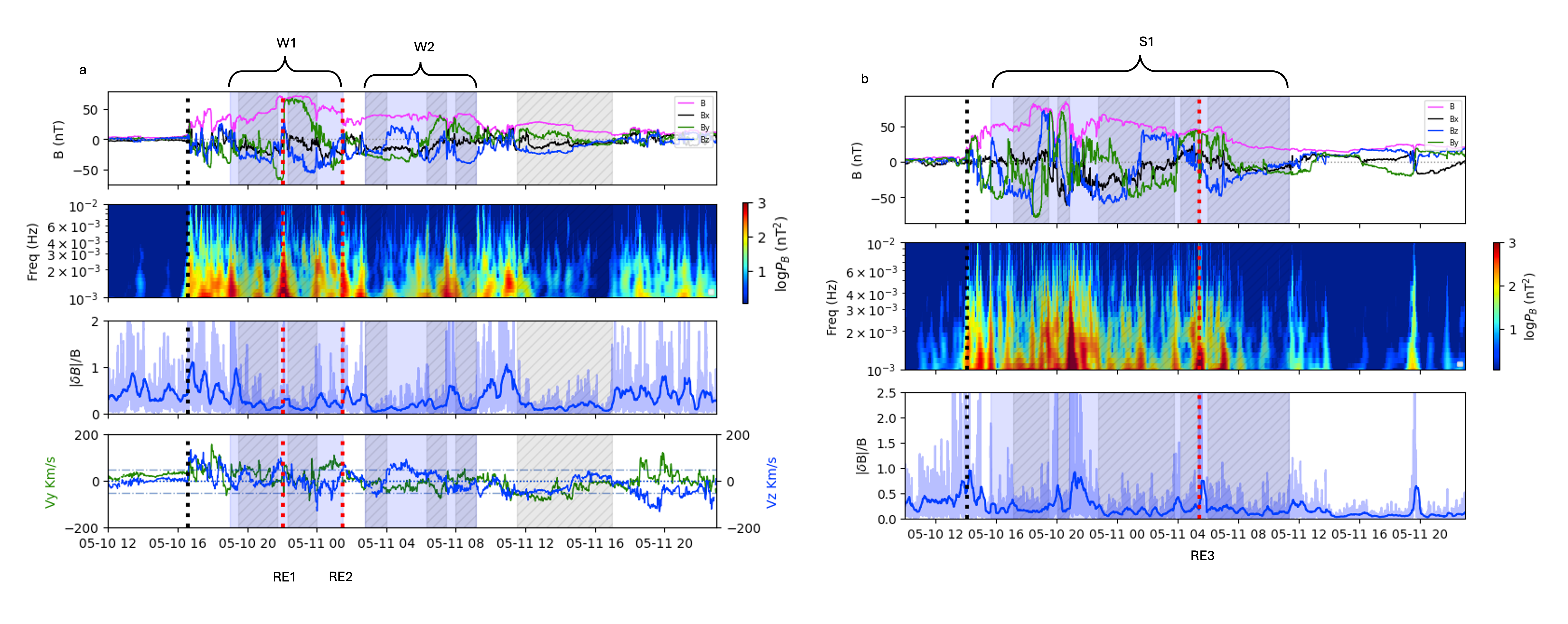}
  \caption{Trace PSD of magnetic field fluctuation $P_B$ in inertial MHD range and fluctuation amplitude $|\delta \mathbf{B}|/B$ from Wind and STA observations, and transverse velocity components $V_{y,z}$ obtained from Wind observation. The horizontal lines in $V_{y,z}$ panel, indicate $V_{y,z}=50$ Km/s. The hatched regions indicate the intervals of ejecta cores. The black and red vertical lines indicate the shock resulted from the arrival of complex solar wind structures and reconnection exhausts (REs, discussed in Section \ref{3.3}), respectively.  }
  \label{fig2}
\end{figure*}

We investigated the trace power spectral density (PSD) of magnetic field fluctuation $P_\mathbf{B}$ (i.e. $\mathbf{B}-\mathbf{B}_{mean}$ fluctuation power) in $1e^{-3} - 1e^{-2}$ Hz range, which corresponds to an inertial range of MHD turbulence. This allows us to analyze the scale of fluctuations not resulting from the rotational field lines of magnetic ejecta. The wavelet spectrogram of the trace PSD $P_\mathbf{B}$ is shown in the second panel of Figure \ref{fig2}a and b obtained from Wind and STA, respectively. The next panel shows normalized fluctuation amplitude $|\delta \mathbf{B}|/B$. The fluctuation amplitude is defined as $|\delta \mathbf{B}|= |\mathbf{B(t)}-\mathbf{B(t+\tau)}|$, where $\tau$ is the timescale of fluctuation, chosen as 94 seconds \citep{2021FrASS...7..109K} corresponds to the inertial range of fluctuation. The intensity of $P_\mathbf{B}$ and $|\delta \mathbf{B}|/B$ usually remains higher within non-ejecta \citep{2023ApJ...956L..30G, 2020ApJS..246...53C} and distorted ejecta than an unperturbed one. This concept was successfully utilized to auto-identification of ejecta in the in situ observations by employing machine learning techniques \citep{2024ApJ...972...94P}. We utilize $P_\mathbf{B}$ and $|\delta \mathbf{B}|/B$ combined with plasma characteristics to confirm the interval of unperturbed ejecta cores within $W1$, $W2$ and $S1$ and show them using hatched intervals in Figure \ref{fig2} and also in Figure\ref{fig1}. The intervals are decided based on low $P_\mathbf{B}$ and $|\delta \mathbf{B}|/B$, coinciding with mostly bi-directional PAD, $N_a/N_p>0.08$, $\beta<0.1$, lower $N_p$, and $T_p<T_{ex}$ intervals. The $N_a/N_p>0.08$ is a good indicator for ejecta cores \citep{Yogesh2022}. With this approach, we identified two separate ejecta cores within $W1$ and three within $W2$, while the $S1$ interval is composed of five distinct cores. 
\par
In the last panel of Figure \ref{fig2}a, we provide the transverse solar wind velocity $V_y$ (in green) and $V_z$ (in blue). We located two highly sheared regions of duration more than an hour with non-radial velocity component $V_{y}>50$ and $V_{z}>50$ km/s within $W1$ and $W2$, respectively. The first sheared region associates with a shock within $W1$, noted by \cite{2025SpWea..2304260W}. A Non-radial velocity component may present at a sheath region following an interplanetary shock and can also manifest Alfv{\'e}nic fluctuations resulting from interaction among field lines \citep{2025ApJ...978..146S}. Furthermore, expansion of the ejecta may lead to the development of non-radial flow within the structures \citep{2022ApJ...927...68A}. However, signatures of over-expansion were not noted during the $W1$ and $W2$ intervals. 
\par 

The solar wind following $W2$ shows signatures of an individual ejecta interval evident from low intensity of $P_\mathbf{B}$ and $|\delta \mathbf{B}|$, $N_a/N_p>0.08$, decrease in $N_p$ and $\beta<0.1$ and $T_p<T_{ex}$. This ejecta does not contain tightly wound magnetic field lines as indicated by the the behavior of magnetic field vectors and  the absence of a localized helical region in the $H_m$ spectrum during the interval. Such characteristics  may occur when a spacecraft crosses through an ejecta leg. \par

\subsection{Reconnection exhausts within complex intervals} \label{3.3}
\begin{figure*}[t]
\centering
    \includegraphics[width=1\textwidth]{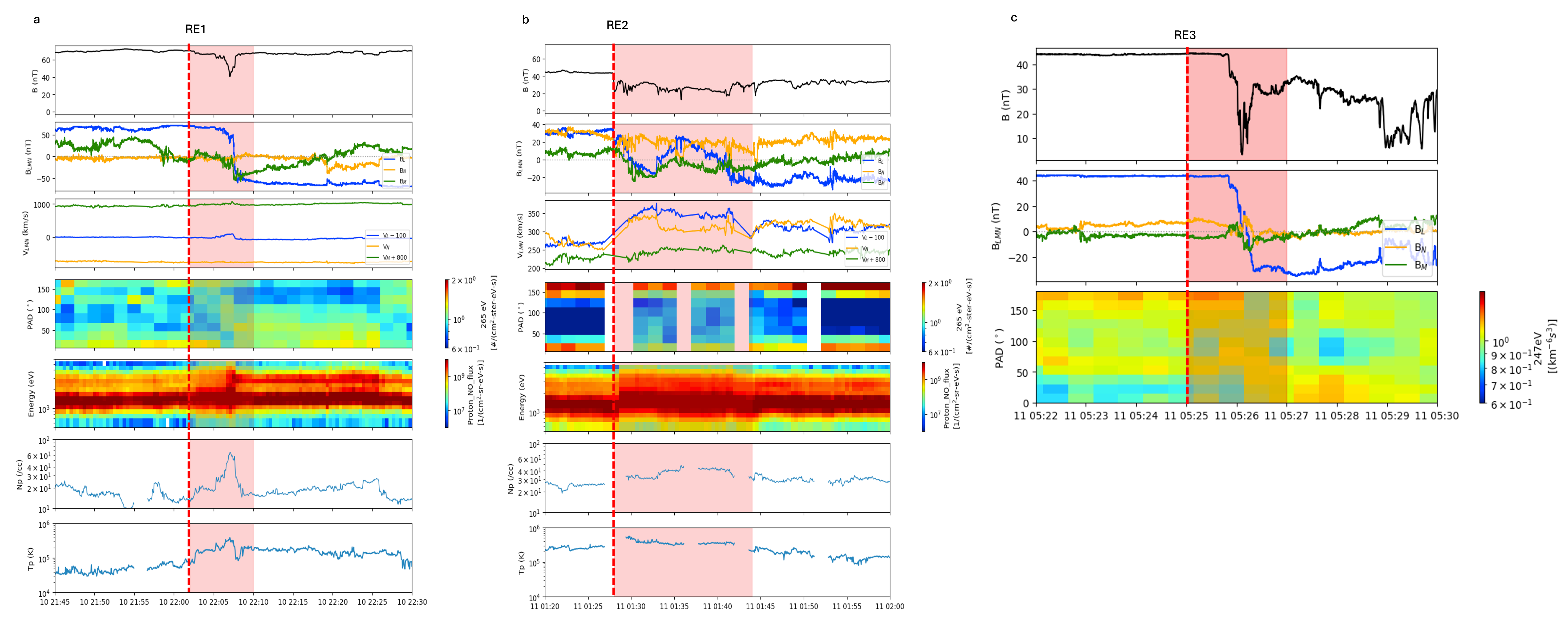}
  \caption{Solar wind magnetic and plasma embedding RE intervals RE1, RE2, and RE3 (in red) observed by MAG, SWE, 3DP instruments onboard Wind (a, b) and MAG, SWEA instruments onboard STA/IMPACT (c), respectively. Note that in a and b, the $V_{L,M,N}$ has been rescaled for clarity.}
  \label{fig3}
\end{figure*}

From the previous Sections, it is clear that the complex intervals $W1, W2$ and $S1$ are formed of  multiple ejecta. Therefore, in this Section, we look for the presence of reconnection exhausts (REs) within the complex intervals where the ejecta cores are not present. We transformed $B_{x,y,z}, V_{x,y,z}$ from GSE to the LMN coordinate ($B_{LMN}, V_{LMN}$) system where L, M, and N indicate the maximum, intermediate, and minimum variance directions, and are obtained by minimum variance analysis \citep[MVA;][]{1968JGR....73.1757S, 2013ApJ...763L..39G}. In this coordinate, L is assumed to point towards the reconnection outflow, M along the reconnection X-line, and N towards the normal direction of the reconnection current sheet \citep{2006Natur.439..175P}. We use suprathermal electron PAD, ion energy spectrum $I_E$ in 500-5000 eV range, $N_p$ and $T_p$ measurements to identify REs because field line connectivity at REs may lead isotropic or bi-directional PAD \citep{2009IAUS..257..367G} and energy conversion to other forms during reconnection may intensify $I_E, N_p$ and $T_p$. In Figure \ref{fig3}, we show intervals of REs in red, and also indicate them in Figures \ref{fig1} and \ref{fig2} using red dotted vertical lines. 
\par
In Wind observation, we located two REs, one within $W1$ and another between $W1$ and $W2$, where $B$ decreases, $B_L$ changes direction, $|V_L|$ has elevated value due to reconnection outflow, $N_p$, $T_p$ show enhancement, PAD becomes isotropic and $I_E$ shows enhanced flux abundance. The limited coverage, lack of plasma velocity vectors and low resolution of plasma data from STA made it difficult to locate REs within $S1$. Using only $B$, $B_{LMN}$, electron PAD and low-resolution $N_p$ and $T_p$ measurements from STA, we confidently located only one RE within $S1$ and show characteristics in Figure \ref{fig3}c. This RE is formed between the last two ejecta cores within $S1$. Nevertheless, $S1$ may contain more REs which were difficult to identify due to STA data insufficiency. \par

\subsection{Complex intervals in the heliospheric imagers} \label{3.4}

\begin{figure*}[t]
  \centering
    \includegraphics[width=0.9\textwidth]{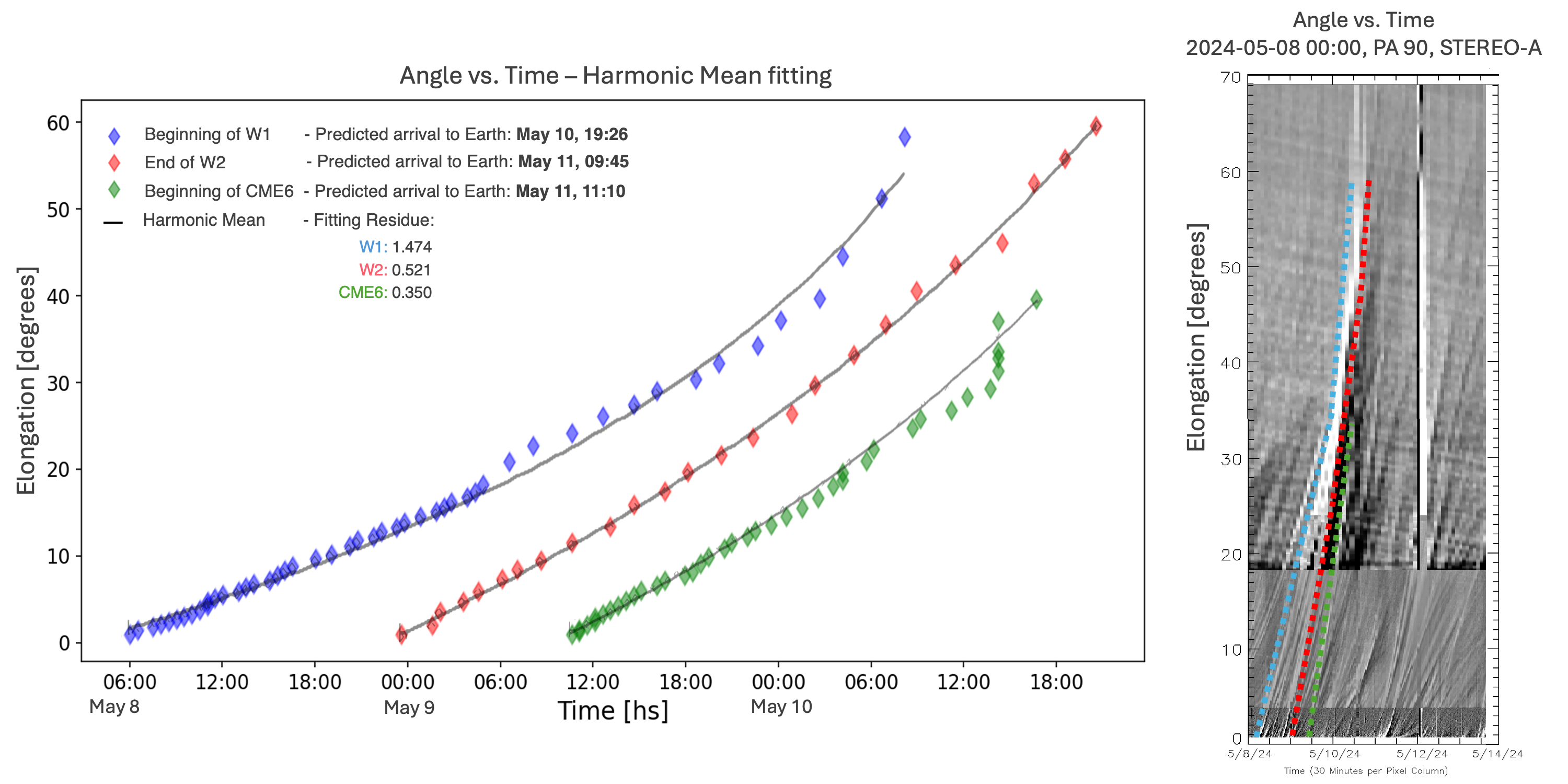}
  \caption{Left panel: Elongation angle vs. time for the three tracked structures on May 8-10: Beginning of W1 (blue diamonds), end of W2 (red diamonds) and CME6 leading edge (green diamonds). The black continuous line denotes the best HM fit for the three structures. Predicted arrivals to Earth and fitting residues are presented in the figure. Right panel: The Jmap at a position angle of 90 $^\circ$ for a combination of COR2, HI1 and HI2 on board STEREO-A. Dashed lines represent the mentioned tracked structures: W1, W2 and CME6 on light-blue, red and green, respectively. 
}
  \label{fig4}
\end{figure*}

Presence of multiple ICMEs in the complex intervals -- $W1$, $W2$ and $S1$, is further investigated using heliospheric imagers H1 and H2 onboard STA. In the right panel of Figure \ref{fig4}, we provide a time-elongation ($\epsilon$) map (J-map) formed of stacking running-difference images of STA/HI1 and HI2 during May 8-13, 2024, within a slit along the ecliptic plane. At the $20^\circ<\epsilon<60^\circ$ in the J-map, we identified a large bright feature formed of multiple CME leading edges. We tracked the leading (in blue symbol) and trailing (in red symbol) edges of the bright feature shown in until $\epsilon=60^\circ$ (Left panel on Figure \ref{fig4}), and fitted them using HM technique to derive their arrivals at the Wind spacecraft located at $\epsilon \sim 98^\circ$. Our estimated arrival of the leading and trailing edges of the large bright feature was within 16 and 10 minutes of the observed start and end times of $W1$ and $W2$, respectively. 
\par
While tracking the large bright feature down to the solar origin up to $\epsilon=2.5^\circ$, we found its association with CMEs (CME1-CME5 in Table \ref{T1}) originating on May 8, including the filament-associated CME3. Another separate bright front observed in the J-map, marked in green, could be unambiguously tracked up to $\epsilon=36^\circ$. This track could be associated with the CME6 in Table \ref{T1}, which originated early on May 9 and had a speed higher than the previous CMEs. In J-map, the leading edge of the CME6 (in green) has been extrapolated until the Earth's location using the HM technique.  It is estimated to arrive at Wind $\sim$20 minutes after the observed start time of the ejecta corresponding to the last hatched interval in Figure \ref{fig1} and \ref{fig2} with a speed higher than preceding events.\par

\section{Discussion and Conclusion}\label{sec4}

The analysis performed in this study of the complex solar wind interval featuring multiple ICMEs on May 10-11, 2024, reveals notable differences between two  heliospheric locations separated by  $\sim 12 ^\circ$ in longitude and 1.5 $^\circ$ in latitude. At Wind, the total magnetic energy and helicity within the 0.5 - 64 hour scale range were found to be 60\% and 35\% than those measured at STA, respectively. Therefore, had part of the complex ICME structure encountered by STA hit the Earth instead, it could have caused even more intense geomagnetic storm than was observed in the actual scenario. However, for a given amount of solar wind energy, resulting geomagnetic disturbance may vary substantially as the solar wind-magnetosphere coupling depends on several factors besides solar wind energy, e.g. solar wind speed, density, transverse component of solar wind magnetic field, clock angle \citep{2022SpWea..2002989L}, preconditioning of the magnetosphere, presence of solar wind structures e.g. ICMEs representing the most energetic and geoeffective driver \citep{2020ApJ...896..149T}. The probable reason behind the significant difference in magnetic content of the Wind and STA counterparts, despite the moderate longitudinal separation of the observing spacecraft is discussed below. \par

The regions $W1$ and $S1'$ contain a complex, merged highly helical structure with enhanced magnetic energy, resulting from the merging of ejecta associated with CME1 and CME2, where the merging process is further evident at the smaller scales in the helicity spectrum. The high magnetic energy is associated with the larger fluctuations in magnetic field lines in the larger scale sizes corresponding to the injection range of MHD frequency. 
 
Both CMEs originated from the same solar source  approximately 7 hours apart and both carried right-handed magnetic field lines. Their flux rope types derived from field line rotational directions and axial tilt mentioned in Table \ref{T1} indicate that their handedness and axial orientations favor magnetic reconnection might have initiated at their early propagation. The evidence for ongoing reconnection is even found from the presence of a reconnection exhaust (RE1) within $W1$ between the first two ejecta cores `WC1' and `WC2'. \par
When fitting the linear force-free flux rope model \citep{1990JGR....9511957L} to three ejecta cores (hatched intervals) within $W2$, we found that the side structures exhibited right-handed helicity, while the central one displayed left-handed helicity. The inclination angles of the flux ropes were $\theta=-7^\circ$, $-14^\circ$ and $-13^\circ$, and azimuthal angle $\phi=308^\circ$, $176^\circ$ and $5^\circ$, respectively. This indicates that $W2$ includes the flank of a left-handed ejecta originating from the filament associated CME3 that propagated at least 27$^\circ$ eastward from the Sun-Earth line \citep{2025SpWea..2304260W}. This ejecta core (`WC3') was bounded by two right-handed ejecta cores `WC4' and `WC5', which is in line with the investigation by \cite{2025SpWea..2304260W}. Using J-map and the ELliptical Evolution model \citep{2015NatCo...6.7135M}, \cite{2025SpWea..2304260W} concluded that the CME3 being overtaken by the following right-handed CME4 sourcing `WC4', and arriving earlier than `WC3' at Wind spacecraft. The CME3 and CME4 may not have magnetically reconnected because of their unfavorable orientation (opposite handedness with non anti-parallel axes). Also, the $H_m$ map indicate that `WC4' and `WC3' may not reconnect as they correspond to $H_m>0$ and $H_m<0$ in the away magnetic field sector indicating anti-clockwise and clockwise rotations, respectively. In Wind in situ observation, a sheared region with $V_z>50$ km/s, comparatively higher $N_a/N_p$, $N_p$, $\beta$ and $P_B$ appeared in between `WC4' and `WC3', indicates a sheath-like structure. The opposite handedness and almost an anti-parallel axial orientations (with an axial separation of $\sim 171 ^\circ$) of `WC3' and `WC5' within $W2$ favors magnetic reconnection. This interpretation is further evidenced from the intensified electron PAD at two opposite directions. However, the CME5 sourcing `WC5' propagated far westward from the CME3 propagation direction with a separation angle of $\sim 65 ^\circ$. The CME5 and CME6 had an initial propagation direction $\sim38^\circ$ and $\sim27^\circ$ westward from the Sun-Earth line, respectively. Therefore, it was hard to track these CMEs at a higher elongation angle in the J-map which was prepared at a position angle of 90$^\circ$. 

\par

Except the left-handed CME3 propagated much eastward $\sim40^\circ$ to STA and far from reaching at STA location, all other CMEs were westward to the Sun-Earth line and had a high speed. Therefore, had a more head-on impact on the STA. All right-handed ICMEs associated with the same polarity inversion line belonging to the bottom part of the AR13664 \citep{2024A&A...692A.112W} had an almost parallel axial orientations. They magnetically reconnected and merged to form a large-complex ejecta. In contrast to Wind observation, at STA, it was hard to individually associate the ejecta cores to their progenitor CMEs. The complex ejecta appeared with a characteristics core size exceeding $ \Delta S>$7 hrs and a strong helical field lines with $\sigma_m=0.9$ at STA.  \par

A detailed study of solar wind energy-Dst distribution performed by \cite{2020ApJ...896..149T} demonstrates a robust correlation between the amount of energy stored in the solar wind interacting with the Earth’s magnetosphere and the intensity of the induced geomagnetic disturbance. For exploring the scatter in Dst for a fixed energy input, contributions of the solar wind bulk speed and $B_z$ intensity were analysed in energy-Dst space. This implies slow solar wind has negligible effect on Earth regardless of its energy content, and solar wind with low and mid energies may induce geomagnetic disturbances by long duration and strong southward $B_z$ driving magnetic reconnection between the solar wind and magnetosphere. Moreover, high-energy and high-speed solar events can result in severe geomagnetic disturbances irrespective of the their field line orientations. The STA's limitation in plasma measurements generate caveats in detail assessment of solar wind-magnetosphere coupling factors of the STA counterpart. The initial propagation properties inferred from the coronagraph and HI analysis in combination with the solar wind in situ magnetic measurement analysis indicate an arrival of a high-speed, high-energy complex solar wind interval carrying multiple interacting ICME ejecta, which further conforms with the hypothesis that the STA counterpart had a higher geoeffectivity than that of the Wind counterpart. However, in a complex solar wind scenario, the hypothesis of a comparatively high energetic solar wind driving a larger geomagnetic response needs to be further investigated considering the influence of other coupling factors which calls for numerical MHD driven magnetospheric simulations, and statistical studies across multiple interacting ICME ejecta.

\par

In this study, we investigate the in situ magnetic and plasma properties of a complex solar wind interval during May 10-11, 2024, when one of the recorded strongest geomagnetic storms occurred, to understand its formation and study its magnetic content that made it capable of driving such a huge storm. With the use of plasma properties and magnetic field fluctuation analysis in the inertial range of frequency, we located six (five) ICME ejecta cores within the complex interval counterparts hitting Wind (STA) spacecraft. Analyzing solar wind magnetic field in time-frequency domain across a large range of frequencies, we quantified the magnetic helicity, energy and the helical core scale-size of the complex ejecta at Wind and STA counterparts and found the reason of their substantial differences even with a moderate angular separation of 12$^\circ$ (0.04 au) in the heliosphere. We attributed these differences primarily to the complicated interaction among ejecta at Wind counterpart, where a left-handed ejecta appeared between two right-handed common-origin ejecta with a non-favorable orientation for strong reconnection. Whereas, in the STA counterpart, ejecta strongly reconnected due to all of them having right-handed rotations and parallel axial orientations and a common origin.

\begin{acknowledgements}
We acknowledge the data source \href{https://cdaweb.gsfc.nasa.gov/istp_public/}{Goddard Space Flight Center's Space Physics Data Facility (SPDF)} from where we downloaded the in situ solar wind data and use of \href{https://www.jhelioviewer.org/}{JHelioviewer}. We thank Wind and STEREO mission teams for making the data publicly available. S.P is thankful for the support of NASA Postdoctoral Program fellowship, L.K.J. thanks for the support of the STEREO mission and Heliophysics Guest Investigator Grant 80NSSC23K0447. T.N-C. thanks for the support of the Solar Orbiter and Parker Solar Probe missions, Heliophysics Guest Investigator Grant 80NSSC23K0447 and the GSFC-Heliophysics Innovation Funds. Y. acknowledge support by NSF grant AGS-2300961 and NASA grant 80NSSC24K0724.

\end{acknowledgements}
\bibliographystyle{aa}
\bibliography{aa}

\end{document}